\documentclass[notitlepage,prd,showpacs,superscriptaddress,nofootinbib,floatfix,showkeys,10pt]{revtex4-1}
\usepackage{graphicx}
\usepackage{amsmath}
\usepackage{bm}
\usepackage{yhmath}
\usepackage{mathtools}
\usepackage{wasysym}
\usepackage{subfigure}
\usepackage{xcolor}
\definecolor{CherryRed}{rgb}{.65,0,.2}
\definecolor{RubyRed}{rgb}{.88,0.07,.3}
\definecolor{CralRed}{rgb}{1,0.25,.25}
\definecolor{CobaltBlue}{rgb}{0,0.28,.67}
\definecolor{RoyalBlue}{rgb}{0.25,0.41,.88}
\definecolor{EmeraldGreen}{rgb}{0.31,0.78,.47}
\definecolor{EmeraldGreen}{rgb}{0.31,0.78,.47}
\definecolor{LimeGreen}{rgb}{50,205,50}
\definecolor{ForestGreen}{RGB}{34,139,34}
\definecolor{PineGreen}{RGB}{1,121,111}\usepackage{cases}
\usepackage[colorlinks=true, linkcolor=EmeraldGreen,citecolor=RoyalBlue,urlcolor=ForestGreen,linkcolor=ForestGreen]{hyperref}
\usepackage{subfigure}
\usepackage{times}
\usepackage{booktabs,bm}
\usepackage{slashed}
\usepackage{amsfonts,amssymb,stmaryrd,latexsym,amsmath}
\usepackage{textcomp}
\usepackage{multirow}
\usepackage{cancel}
\usepackage{array}
\usepackage{hyperref}
\usepackage{pgf}
\usepackage{orcidlink}

\def\SU3{{\text{SU(3)}_{\rm F}}}

\def \pcs4338{{P_{\psi s}^\Lambda(4338)^0}}

\begin{document}
	
	\title{\textcolor{CobaltBlue}{Masses of Purely Top-Quark Bound States: Toponium and the Triply-Top Baryon }}

	\author{Z.~Rajabi Najjar\orcidlink{0009-0002-2690-334X}}
	\email{rajabinajar8361@ut.ac.ir }
	\affiliation{Department of Physics, University of Tehran, North Karegar Avenue, Tehran 14395-547,  Iran }

	\author{K.~ Azizi\orcidlink{0000-0003-3741-2167}}
	\email{kazem.azizi@ut.ac.ir}
	\thanks{Corresponding author}
	\affiliation{Department of Physics, University of Tehran, North Karegar Avenue, Tehran 14395-547,  Iran }
	\affiliation{Department of Physics,  Dogus University, Dudullu-\"{U}mraniye, 34775 Istanbul, T\"urkiye }

	\date{\today}

	\begin{abstract}
      We investigate the pseudoscalar ($\eta_t$) and vector ($\psi_t$) toponium states, as well as the triply-top baryon ($\Omega_{ttt}$), using the QCD sum-rule method. This study was motivated by the recent observation of a pseudoscalar enhancement near the $t\bar{t}$ threshold, reported by the CMS and ATLAS collaborations with a statistical significance exceeding $5\sigma$. In the calculations, we consider both the perturbative and nonperturbative contributions, with the nonperturbative operators taken into account up to dimension eight. The results obtained for the pseudoscalar toponium provide a theoretical estimate that is consistent with the near-threshold events observed in recent experimental studies. The calculated  negative binding energy for both the pseudoscalar and vector toponium states reflects the strong correlation within the $t\bar{t}$ system and can be interpreted as $t\bar{t}$ bound states, while the calculated central mass for the $\Omega_{ttt}$ slightly exceeds the central value of the sum of the constituent top-quark masses. The results of this study can provide a precise theoretical guide for future experimental investigations of these states, which are composed entirely of top quarks, at high-energy colliders such as the LHC and future facilities like the FCC.
                        	\end{abstract}

	\maketitle
	\renewcommand{\thefootnote}{\#\arabic{footnote}}
	\setcounter{footnote}{0}


\section {Introduction}\label{sec:one}
The top quark, the heaviest known elementary particle (discovered in 1995 at the Tevatron accelerator at Fermilab by the CDF and D$\O$  experiments \cite{CDF:1995wbb,D0:1995jca}), plays a decisive role in contemporary particle physics—not only because of its exceptional mass but also due to its unique dynamical properties. The significance of this particle goes beyond serving as a precise test of the Standard Model (SM); indeed, the top quark provides direct access to research domains beyond the Standard Model (BSM). However, its most striking characteristic, which distinguishes it from all other quarks, is its exceptionally short lifetime (about $ 5 \times 10^{-25}$ s), shorter than the timescale required for hadronization \cite{Bigi:1986jk,Bernreuther:2008ju}. This short lifetime has traditionally been cited as the basis for the argument that bound states such as toponium—a quark–antiquark $t\bar{t}$ system—cannot form. This assumption has even been stated in fundamental reference works on particle physics. It is this paradoxical combination, an enormous mass together with an extraordinarily short lifetime, that has made the top quark one of the central subjects of investigation in the SM and BSM research domains over the past few decades \cite{Schrempp:1996fb,Bhat:1998cd,Cvetic:1997eb,Atwood:2000tu,Plehn:2011tg,Chakraborty:2003iw,Merkel:2004td,Wagner:2005jh,Quadt:2006dqn,Kats:2009bv,Galtieri:2011yd,Palle:2012mr,Schilling:2012dx,Hagiwara:2016rdv}.

Recent experimental findings have called into question the long-standing assumption that bound states of the top quark cannot form. In particular, the CMS \cite{CMS:2025kzt} and ATLAS \cite{ATLAS:2025mvr} Collaborations at the LHC reported a statistically significant excess near the $t\bar{t}$ threshold, consistent with the production of a pseudoscalar quasi-bound toponium state.  A key quantum signature of top–anti-top systems near threshold, even at the highest accessible mass scales, is the strong spin correlation between the top and anti-top quarks, which may reflect quasi-bound dynamics or significant quantum entanglement. Such entanglement, although not directly observed, may influence the behavior of the $t\bar{t}$ system near threshold, particularly given the extremely short lifetimes of its constituents. This experimental observation, achieved with a statistical significance exceeding $5\sigma$ \cite{CMS:2025kzt} through the observation of its decay products, rapidly motivated a wide range of theoretical investigations \cite{Fuks:2025toq,Matsuoka:2025jgm,Sjostrand:2025qez,Fuks:2025wtq,Goncalves:2025hyx,Zhu:2025ezg,Zhang:2025fdp,Lopez:2025kog,Thompson:2025cgp,Zhang:2025xxd,Shao:2025dzw,Bai:2025buy,Ellis:2025nkm,LeYaouanc:2025mpk,Xiong:2025iwg,Fu:2025yft,Fu:2025zxb,Fuks:2025sxu,Afik:2025ejh,Luo:2025psq}.  It is noteworthy that numerous theoretical predictions  had previously anticipated the existence and properties of toponium \cite{Fadin:1987wz,Kuhn:1987ty,Barger:1987xg,Strassler:1990nw,Fadin:1990wx,Fadin:1994pj,Hoang:2000yr,Hagiwara:2008df,Penin:2005eu,Sumino:2010bv,Kiyo:2008bv,Beneke:2015kwa,Kawabata:2016aya,Reuter:2018rbq,Fuks:2021xje,Aguilar-Saavedra:2024mnm,Jiang:2024fyw,Jafari:2025rmm,Francener:2025tor,dEnterria:2025ecx,Garzelli:2024uhe,Wang:2024hzd,Fuks:2024yjj}.

Based on recent findings concerning toponium, it has been suggested that baryons containing three top quarks, $\Omega_{ttt}$,  might also exist within the framework of the SM.   This hypothetical triply top baryon, expected to be the heaviest baryon known, would provide a unique laboratory for probing the behavior of the strong interaction at unprecedented mass scales. Although the extremely short lifetime of the top quark (with a decay width of $\Gamma_t = 1.3~\text{GeV}$ \cite{Chen:2023osm,Chen:2023dsi}) has traditionally been regarded as an obstacle to hadronization, recent analyses indicate that the timescale associated with the formation of the $\Omega_{ttt}$ and its corresponding Bohr radius might still allow the emergence of a bound state within the top-quark lifetime \cite{Fu:2025yft,Fu:2025zxb,Xiong:2025iwg,Jia:2006gw}. Consequently, the systematic investigation of the triply top baryon can be pursued using theoretical frameworks and tools previously developed for triply heavy baryons, such as $bbb$ and $ccc$. Over the past several decades, a large body of studies on the spectroscopy, production, and decay properties of  triply heavy baryons has employed a wide range of theoretical approaches, including potential models, sum rules, and lattice QCD simulations. As a result, although an extensive body of theoretical work exists for triply heavy baryons, the formation and experimental identification of the $\Omega_{ttt}$—owing to challenges exceeding those of toponium and the current lack of direct empirical evidence—require a dedicated reassessment and adaptation of these theoretical frameworks, representing a critical frontier test of the SM.

In this work, we perform a systematic analysis of the masses of bound states in the toponium system, including the pseudoscalar state, $\eta_t$, and the vector state, $\psi_t$, as well as the mass of the triply top baryon, $\Omega_{ttt}$, within the framework of QCD sum rules. This approach, based on the QCD Lagrangian, provides a powerful tool for predicting hadronic properties, particularly in heavy quark systems, and its efficacy has been previously demonstrated in successful predictions of heavy hadron properties in agreement with experimental data \cite{Aliev:2009jt,Aliev:2010uy,Aliev:2012ru,Agaev:2016mjb,Azizi:2016dhy}.
Mass calculations within this framework require the use of a two-point correlation functions constructed from suitable interpolating currents, wherein the Operator Product Expansion (OPE), Borel transformation, and continuum subtraction based on quark–hadron duality are employed. In the present study, the OPE is considered up to dimension-eight non-perturbative condensates.
The results of this analysis are expected to offer valuable theoretical insights for the interpretation of forthcoming LHC data and can inform future collider studies, such as those planned for CEPC and FCC-ee. The structure of the paper is as follows: Section \ref{sec:two}   introduces the QCD sum-rule formalism and computational details; Section \ref{sec:three}   discusses and compares the numerical results for the masses of $\eta_t$, $\psi_t$, and $\Omega_{ttt}$; finally, Section \ref{sec:four} presents a summary and conclusions of this work. Alternatively: Relevant expressions are  included in the Appendix.

\section {QCD sum rule Formalism for the full-top systems}\label{sec:two}

In this study, the QCD sum rules framework is employed to determine the masses of the pseudoscalar,  and vector toponium states, as well as that of the triply top baryon. This approach provides a powerful and systematic method that connects hadronic parameters to the fundamental degrees of freedom of quarks and gluons within QCD. The present analysis is divided into two main parts: i) the physical side, where the correlation function is rewritten in terms of hadronic parameters such as the mass and decay constant; and ii) the QCD side, where the same function is expanded through the OPE, including the contributions from  the gluon condensates. By constructing these two representations of a single correlation function, we relate them using the dispersion relation and the assumption of quark–hadron duality, thereby formulating the QCD sum rules for the physical quantities of interest. Finally, to extract the desired parameters, the coefficients of the corresponding Lorentz structures obtained from each side must be matched consistently. In this regard, to derive the sum rules corresponding to the toponium states (pseudoscalar and vector) and the triply top baryon ($\Omega_{ttt}$), it is essential to define an appropriate correlation function. These functions, constructed from the time-ordered product of hadronic currents, forms the starting point for calculations within the QCD sum rule framework and is defined as follows:
\begin{eqnarray}
\Pi^{PS} &=&i\int d^{4}xe^{iqx}\langle 0|\mathcal{T}\{\eta^{PS} (x)  {\eta}^{\dagger PS}(0)\}|0\rangle,\nonumber\\	
\Pi_{\mu\nu}^{V}(q)&=&i\int d^{4}xe^{iqx}\langle 0|\mathcal{T}\{\eta_\mu^{V} (x)\eta_\nu^{\dagger V} (0)\}|0\rangle,\nonumber\\		\Pi_{\mu\nu}^{B}(q)&=&i\int d^{4}xe^{iqx}\langle 0|\mathcal{T}\{\eta_\mu^{B} (x)\bar{\eta}_\nu^{B} (0)\}|0\rangle. 
		 \label{eq:CF1}
	\end{eqnarray}
	In these relations, $\eta^{PS}$, $\eta^{V}$ and $\eta^{B}$ represent the interpolating currents for $\eta_t$, $\psi_t$ and $\Omega_{ttt}$, respectively. $\mathcal{T}$ denotes the time-ordered product and $q$ stands for the four-momentum of the hadron. To derive the QCD sum rules corresponding to the states under investigation, the appropriate interpolating currents for each system are defined as follows:
	\begin{eqnarray}
\eta^{PS} (x)&=&\overline{t}^{ a}(x)\gamma_5t^{ b}(x),\nonumber\\	
\eta^{V}_\mu (x)&=&\overline{t}^{ a}(x)\gamma_\mu t^{ b}(x),\nonumber\\	
\eta^{B}_\mu(x) &=&  \epsilon^{abc} (t^{aT} C \gamma_\mu t^{ b}) t^c.
	\label{cur1}
	\end{eqnarray}	
	In the above expressions for the hadronic currents, $C$ denotes the charge conjugation operator, and the indices $a$, $b$ and $c$ represent color indices. In the next step, the correlation function constructed from these currents is analyzed separately on the physical (phenomenological) side and the QCD (OPE) side.	
 \subsection{Physical side} 	
Within the framework of the sum rules, to obtain the physical (phenomenological) part of the correlation function, a complete set of intermediate hadronic states with quantum numbers matching those of the interpolating current  for each system is inserted into the calculation. By performing the four-dimensional integral over spacetime coordinates and isolating the ground-state contribution from excited states, the correlation function for the pseudoscalar toponium, $\eta_t$, can be expressed in its hadronic form as follows:
	\begin{eqnarray}
\Pi^{\mathrm{Phys}}&=&\frac{{\langle}0| \eta^{PS} | \eta_t(q,s)\rangle \langle \eta_t(q,s)| \eta^{\dagger PS}|
 0\rangle}{m_{PS}^2-q^2}+\cdots.	
	\label{Eq:cor:Phys1}
	\end{eqnarray}	
	Also, for the vector toponium, $\psi_t$:
	\begin{eqnarray}	
\Pi^{\mathrm{Phys}}_{\mu\nu}&=&\frac{{\langle}0| \eta^{V}_\mu | \psi_t(q,s)\rangle \langle \psi_t(q,s)| \eta^{\dagger V}_\nu|
 0\rangle}{m_{V}^2-q^2}+\cdots.
 	\label{Eq:cor:Phys2}
	\end{eqnarray}		
	Similarly, for  the triply top baryon, $\Omega_{ttt}$, the correlation function is expressed as follows:
	\begin{eqnarray}
\Pi^{\mathrm{Phys}}_{\mu \nu}(q)&=&\frac{\langle0|\eta^{B}_{\mu}| \Omega_{ttt}(q,s)\rangle\langle \Omega_{ttt}(q,s)|\bar{\eta}^{B}_{\nu}|0\rangle}{m^2_B-q^2}+\cdots.
	\label{Eq:cor:Phys3}
	\end{eqnarray}		
	In the set of Eqs. (\ref{Eq:cor:Phys1}--\ref{Eq:cor:Phys3}), the symbol $\mathbf{\cdots}$ represents the contributions from the excited and continuum states, and $m_{PS}$, $m_V$, and $m_B$ denote the masses of the $\eta_t$, $\psi_t$, and the $\Omega_{ttt}$, respectively. To proceed with the calculations, it is necessary to determine the matrix elements of the interpolating currents between the vacuum and the corresponding hadronic states (mesonic or baryonic). These matrix elements are parameterized in terms of the decay and coupling constants ($f_{PS}$, $f_{V}$, $\lambda$), as well as the relevant masses, and are defined as follows:
	
	\begin{eqnarray}
{\langle}0| \eta^{PS} | \eta_t(q,s)\rangle&=& f_{PS} ~ \frac{m^2_{PS}}{2m_t},\nonumber\\	
{\langle}0| \eta^{V}_\mu | \psi_t(q,s)\rangle&=& f_{V}  ~ m_{V} \varepsilon_\mu^{(\lambda)},\nonumber\\	
\langle 0|\eta_{\mu}^B|\Omega^*_{ttt}(q,s)\rangle&=&\lambda~ u_{\mu}(q,s).
	\label{Eq:Matrixelm}
	\end{eqnarray}		
	
In this context, $\varepsilon_{\mu}$ denotes the four–polarization vector of the $\psi_t$, while $u_{\mu}(q,s)$ represents the Rarita-Schwinger spinor corresponding to the spin-3/2 triply top baryon with spin projection $s$.	By performing the summation over the polarization vectors of the vector meson, we have:
	\begin{eqnarray}\label{sum}
\sum_{\lambda}\varepsilon^{(\lambda)^*}_{\mu}\varepsilon^{(\lambda)}_{\nu}=-(g^{\mu\nu}-q_{\mu}q_{\nu}/m_{V}^2),
\end{eqnarray}
and similarly, summing over the Rarita–Schwinger spinors associated with the baryonic state:	
\begin{eqnarray}
		\sum u_\mu(q,s) \bar{u}_\nu (q,s) &=& -(\!\not\!{q} + m) \Bigg( g_{\mu\nu} - {1
\over 3} \gamma_\mu \gamma_\nu - {2 q_\mu q_\nu \over 3 m_B^2} + {q_\mu
\gamma_\nu - q_\nu \gamma_\mu \over 3 m_B} \Bigg).
				\label{Eq:Summation}
	\end{eqnarray}	
	The final expressions for the physical (phenomenological) parts of the correlation functions for the $\eta_t$, $\psi_t$, and  $\Omega_{ttt}$ are obtained as follows:
\begin{eqnarray}\label{phen2}
\Pi^{\mathrm{Phys(PS)}}&=&\frac{f_{PS}^2 (\frac{m_{PS}^2}{2m_t})^2}{m_{PS}^2-q^2} + \cdots,\nonumber\\
\Pi_{\mu\nu}^{\mathrm{Phys(V)}}&=&\frac{f_{V}^2 m_{V}^2}{m_{V}^2-q^2}\left[-g_{\mu\nu}+\frac{q_{\mu}q_{\nu}}{m_{V}^2}\right] + \cdots,\nonumber\\
\Pi_{\mu\nu}^{\mathrm{Phys(B)}}(q)&=&\frac{\lambda^{2}}{m_{B}^{2}-q^{2}}(\!\not\!{q} + m_{B})\Big[g_{\mu\nu} -\frac{1}{3} \gamma_{\mu} \gamma_{\nu} - \frac{2q_{\mu}q_{\nu}}{3m_{B}^{2}} +\frac{q_{\mu}\gamma_{\nu}-q_{\nu}\gamma_{\mu}}{3m_{B}} \Big]+\cdots.
\end{eqnarray}	
It is worth noting that the interpolating currents of spin-3/2 baryons (such as $\Omega_{ttt}$) can couple not only to spin-3/2 states but also to spin-1/2 baryons with the same quark content. To effectively suppress these unwanted contributions and ensure that only the structures associated with the spin-3/2 states are retained in the analysis, It is essential to identify Lorentz structures that couple solely to the spin-3/2 states (for further details, see \cite{Najjar:2025dzl}).
Accordingly, the final hadronic representation of the correlation function for $\Omega_{ttt}$ can be expressed as follows:
\begin{eqnarray}
\Pi _{\mu \nu}^{\mathrm{Phys(B)}}(q)&=&\frac{\lambda{}^2}{m_B^{2}-q^{2}}  (\!\not\!{q} + m_B)g_{\mu\nu}+\cdots,
\label{eq:CorFun1}
\end{eqnarray}
From Eqs. (\ref{phen2}) and (\ref{eq:CorFun1}), it is evident that the correlation function for the vector toponium contains two independent Lorentz structures, $g_{\mu\nu}$ and $q_{\mu}q_{\nu}$, whereas for the baryon $\Omega_{ttt}$, only the structures $g_{\mu\nu}$ and $\!\not\!{q} g_{\mu\nu}$ are free from spin-1/2 contamination and arise purely from the spin-3/2 component.	
	Finally, after applying the Borel transformation with respect to $q^2$ to eliminate the contributions from higher resonances and the continuum, the hadronic representation of the correlation function for the studied particles is obtained as follows:
\begin{eqnarray}
\mathcal{\widehat B}\Pi ^{\mathrm{Phys(PS)}}(q)&=&f_{PS}^2 (\frac{m_{PS}^2}{2m_t})^2 e^{-\frac{m^{2}_{PS}}{M^{2}}} + \cdots,\nonumber\\
\mathcal{\widehat B}\Pi _{\mu \nu}^{\mathrm{Phys(V)}}(q)&=&f_{V}^2 {m_{V}^2} e^{-\frac{m^{2}_V}{M^{2}}} \left[-g_{\mu\nu}+\frac{q_{\mu}q_{\nu}}{m_{V}^2}\right] + \cdots,\nonumber\\
\mathcal{\widehat B}\Pi _{\mu \nu}^{\mathrm{Phys(B)}}(q)&=&\lambda^2 e^{-\frac{m^{2}_B}{M^{2}}} (\!\not\!{q}+ m) g_{\mu\nu}+\cdots.
\label{eq:CorFunBorel}
\end{eqnarray}	
 \subsection{QCD side}	
In this part, the two-point correlation function is evaluated in the Euclidean region for large spacelike values of the four-momentum ($q^{2} \ll 0$).	The interpolating currents defined in Eq.~(\ref{cur1}
) are substituted into Eq.~(\ref{eq:CF1}), and all possible contractions among the quark fields are performed using Wick’s theorem.Consequently, the correlation function can be expressed in terms of the heavy top-quark propagators, which, for the $\eta_t$, $\psi_t$ and $\Omega_{ttt}$ states, are given as follows:
\begin{eqnarray}
\label{QCDSide1}
\Pi^{\mathrm{PS}} (q) &=& i \int d^4x
e^{iqx}    Tr \Big[\gamma_5
S_{t}^{ a a'} \gamma_5 S_t^{bb'}\Big], \nonumber \\
\Pi_{\mu\nu}^{\mathrm{V}} (q) &=& i  \int d^4x
e^{iqx}   Tr \Big[\gamma_\mu
S_{t}^{ a a'} \gamma_\nu S_t^{bb'}\Big], \nonumber \\
\Pi_{\mu\nu}^{\mathrm{B}} (q) &=& i\,
\epsilon^{abc} \epsilon^{a'b'c'} \int d^4x
e^{iqx}  \Big\{ S^{ca'}_{t}(x)\gamma_{\nu} S'^{ab'}_{t}(x)\gamma_{\mu}S^{bc'}_{t}(x) \nonumber\\
&-&S^{ca'}_{t}(x)\gamma_{\nu} S'^{bb'}_{t}(x)\gamma_{\mu}S^{ac'}_{t}(x)-S^{cb'}_{t}(x)\gamma_{\nu} S'^{aa'}_{t}(x)\gamma_{\mu}S^{bc'}_{t}(x)\nonumber\\
&+&S^{cb'}_{t}(x)\gamma_{\nu} S'^{ba'}_{t}(x)\gamma_{\mu}S^{ac'}_{t}(x)-S^{cc'}_{t}(x)Tr\left[ S^{ba'}_{t}(x)\gamma_{\nu} S'^{ab'}_{t}(x)\gamma_{\mu}\right] \nonumber\\
&+&\left. S^{cc'}_{t}(x)Tr\left[ S^{bb'}_{t}(x)\gamma_{\nu} S'^{aa'}_{t}(x)\gamma_{\mu}\right] \right\rbrace.
\end{eqnarray}	
Here, $S'_t$ is defined through the charge–conjugation operator $C$ as follows:	
	\begin{eqnarray}\label{eq1}
	S'_t&=& C S_Q^T C.
		\end{eqnarray}	
In these calculations, the full top-quark propagator, $S_t(x)$, plays a crucial role, as it incorporates both perturbative and nonperturbative components. The explicit form of this propagator in the coordinate space, following \cite{Agaev:2020zad}, is given as follows:	\begin{eqnarray}\label{eqQCD2}
		&&S_{Q}^{ab}(x)=i\int \frac{d^{4}k}{(2\pi )^{4}}e^{-ikx}\Bigg \{\frac{\delta
			_{ab}\left( {\slashed k}+m_{Q}\right) }{k^{2}-m_{Q}^{2}}-\frac{%
			g_{s}G_{ab}^{\alpha \beta }}{4}\frac{\sigma _{\alpha \beta }\left( {\slashed %
				k}+m_{Q}\right) +\left( {\slashed k}+m_{Q}\right) \sigma _{\alpha \beta }}{%
			(k^{2}-m_{Q}^{2})^{2}}  \notag  \\
		&&+\frac{g_{s}^{2}G^{2}}{12}\delta _{ab}m_{Q}\frac{k^{2}+m_{Q}{\slashed k}}{%
			(k^{2}-m_{Q}^{2})^{4}}+\frac{g_{s}^{3}G^{3}}{48}\delta _{ab}\frac{\left( {%
				\slashed k}+m_{Q}\right) }{(k^{2}-m_{Q}^{2})^{6}}\left[ {\slashed k}\left(
		k^{2}-3m_{Q}^{2}\right) +2m_{Q}\left( 2k^{2}-m_{Q}^{2}\right) \right] \left(
		{\slashed k}+m_{Q}\right) +\cdots \Bigg \}.  \notag \\
		&&
	\end{eqnarray} 
	The propagator $S_t$ contains the gluon field-strength tensor $G^{\alpha \beta}$, which is defined in terms of the Gell-Mann matrices $\lambda^A$ through the following relation:
\begin{equation}\label{eqQCD3}
		G_{ab}^{\alpha \beta }=G_{A}^{\alpha \beta }\lambda_{ab}^{A}/2,\,\,~~t^A=\lambda^A/2.
	\end{equation}	
The quantities $G^2$ and $G^3$ denote the dimension-four and dimension-six gluon condensates, respectively, and are defined as:		
	\begin{equation}\label{eqQCD4}
		G^{2}=G_{\alpha \beta}^{A}G_{A }^{\alpha \beta},\ \ G^{3}=\,\,f^{ABC}G^{\alpha \beta }_{A}G^{\beta
			\delta }_{B}G^{\delta \alpha  }_{C}.
	\end{equation}	
	In these relations, $\alpha$, $\beta$, $\delta$ denote Lorentz indices, and $f^{ABC}$ are the structure constants of the color group $\text{SU}_c(3)$, where $A,B,C=1,\,2\,\ldots 8$. In this analysis, the OPE is considered up to operators of dimension-eight. This expansion includes nonperturbative contributions associated with QCD vacuum condensates, namely $\langle G^2 \rangle$ (dimension-four), $\langle G^3 \rangle$ (dimension-six), and $\langle G^2 \rangle^2$ (dimension-eight). Each term in the QCD side—containing both perturbative and nonperturbative contributions up to dimension eight—has been explicitly evaluated in the present work. For specific technical details related to the expansion in terms of Feynman diagrams and standard QCD sum-rule procedures, we refer the reader to Ref \cite{Najjar:2024deh}.  The final results are expressed as Fourier integrals and can be evaluated using standard mathematical techniques \cite{Najjar:2024deh, Najjar:2024ngm}. As mentioned earlier, on both the QCD and hadronic sides, the $\psi_t$ and $\Omega_{ttt}$ each possess two independent Lorentz structures. These structures are labelled as ($g_{\mu\nu}$, $q_{\mu}q_{\nu}$) and ($g_{\mu\nu}$, $\!\not\!q g_{\mu\nu}$), respectively. In the present work, the structure $g_{\mu\nu}$ is chosen for the vector toponium, and the corresponding structure $\!\not\!q g_{\mu\nu}$ is selected for $\Omega_{ttt}$. After applying the Borel transformation and subtracting the continuum, the final QCD-side expression of the correlation function for the studied particles is given as follows:\begin{eqnarray}
	\Pi^{\mathrm{QCD(PS)}}(s_0,M^2)&=&\int_{(2m_t)^2}^{s_0}ds\,e^{-\frac{s}{M^2}}\rho(s)+\Gamma(M^2),\notag \\
		\Pi^{\mathrm{QCD(V)}}_{ g_{\mu\nu}}(s_0,M^2)&=&\int_{(2m_t)^2}^{s_0}ds\,e^{-\frac{s}{M^2}}\rho_{ g_{\mu\nu}}(s)+\Gamma_{ g_{\mu\nu}}(M^2),\notag \\
		\Pi^{\mathrm{QCD(B)}}_{\not\!q g_{\mu\nu}}(s_0,M^2)&=&\int_{(3m_t)^2}^{s_0}ds\,e^{-\frac{s}{M^2}}\rho_{\not\!q g_{\mu\nu}}(s)+\Gamma_{\not\!q g_{\mu\nu}}(M^2).		\label{Eq:finalCor:QCD}
	\end{eqnarray}	
	Here, the continuum threshold ($s_0$) separates the contributions of the excited and continuum states from that of the ground state. The spectral densities $\rho _{i}(s)$, which include only the perturbative contributions, are defined as $\rho _{i}(s)=\frac{1}{\pi}\mathrm{Im}[\Pi_{i}^{\mathrm{QCD}}]$.  Furthermore, the functions $\Gamma_i(M^2)$, which have no imaginary parts, encompass the complete set of nonperturbative terms associated with the gluon condensates of mass dimensions four, six, and eight. The precise mathematical expressions for the spectral density $\rho(s)$ and the nonperturbative component $\Gamma(M^2)$, corresponding to the $\eta_t$, are provided, as examples, in the Appendix.
	By equating the representations of the correlation function on the QCD side [Eq.~(\ref{Eq:finalCor:QCD})] and the hadronic side [Eq.~(\ref{eq:CorFunBorel})], and by matching the coefficients of the corresponding Lorentz structures on both sides, the QCD sum rules for the masses of the studied states are obtained as follows:
	\begin{eqnarray}
	f_{PS}^2 (\frac{m_{PS}^2}{2m_t})^2 e^{-\frac{m^{2}_{PS}}{M^{2}}}&=&\Pi^{\mathrm{QCD(PS)}}(s_0,M^2),\nonumber\\
	f_{V}^2 {m_{V}^2} e^{-\frac{m^{2}_V}{M^{2}}} &=&\Pi^{\mathrm{QCD(V)}}_{ g_{\mu\nu}}(s_0,M^2),	\nonumber\\			\lambda^2 e^{-\frac{m^2_B}{M^2}}&=&\Pi^{\mathrm{QCD(B)}}_{~\!\not\!{q} ~g_{\mu\nu}}(s_0,M^2).
	\label{Eq:cor:match1}
\end{eqnarray}
	In the next step of the calculations, the relations corresponding to the masses of the studied states are obtained by differentiating Eq. (\ref{Eq:cor:match1}) with respect to $ -\frac1{M^2} $:
	\begin{eqnarray}
	m^2_{\eta_t}=\frac{\frac{d}{d(-\frac{1}{M^2})}\Pi^{\mathrm{QCD(PS)}}(s_0,M^2)}{\Pi^{\mathrm{QCD(PS)}}(s_0,M^2)},\notag \\ \notag \\
		m^2_{\psi_t}=\frac{\frac{d}{d(-\frac{1}{M^2})}\Pi^{\mathrm{QCD(V)}}_{g_{\mu\nu}}(s_0,M^2)}{\Pi^{\mathrm{QCD(V)}}_{g_{\mu\nu}}(s_0,M^2)},\notag \\	\notag \\			m^2_{\Omega_{ttt}}=\frac{\frac{d}{d(-\frac{1}{M^2})}\Pi^{\mathrm{QCD(B)}}_{~\not\! q~ g_{\mu\nu}}(s_0,M^2)}{\Pi^{\mathrm{QCD(B)}}_{~\not\!q ~g_{\mu\nu}}(s_0,M^2)}.
		\label{Eq:mass:Groundstates1}
	\end{eqnarray}
	In the following section, we present the numerical results of the QCD sum rules for the masses of the considered states, including the pseudoscalar and vector toponium, as well as the triply top  baryon.
\section {QCD Sum-Rule Mass Analysis}\label{sec:three}	
In this section, we present the numerical analysis of the studied states, including the pseudoscalar toponium ($\eta_{t}$), the vector toponium ($\psi_{t}$), and the  triply top  baryon ($\Omega_{ttt}$).	The input parameters used in the calculations are summarized in Table \ref{tab:Parameter}.	As indicated, nonperturbative condensates are considered up to mass dimension-eight. To include contributions from higher-dimensional operators in the OPE, the factorization hypothesis is employed. This approximation expresses higher-dimensional operators as products of lower-dimensional ones, effectively incorporating contributions from operators with $D>6$ into the analysis.	The QCD sum-rule method for extracting hadron masses depends not only on physical input parameters but also on two auxiliary parameters: the Borel parameter $M^2$ and the continuum threshold $s_0$. Proper choice of these parameters is essential to ensure the stability and reliability of the numerical results. Minimal sensitivity of the extracted results to these auxiliary parameters is a standard requirement of the QCD sum-rule approach.	
\begin{table}[htb]
		\begin{tabular}{|c|c|}
			\hline\hline
Quantities&Measurements \\ \hline\hline
									$m_{t}$                                     & $172.56{}^{+0.31}_{-0.31}~\mathrm{GeV}$ \cite{ParticleDataGroup:2022pth}\\ \hline
					$\langle \frac{\alpha_s}{\pi} G^2 \rangle $ & $0.012{}^{+0.004}_{-0.004}$ $~\mathrm{GeV}^4 $\cite{Belyaev:1982cd}\\ \hline
			$\langle g_s^3 G^3 \rangle $                & $ 0.57{}^{+0.29}_{-0.29}$ $~\mathrm{GeV}^6 $\cite{Narison:2015nxh}\\ 
			\hline\hline
		\end{tabular}
		\caption{Summary of the input parameters utilized in the study.}
		\label{tab:Parameter}
	\end{table}

\begin{figure}[h!]
	\begin{center}
		\includegraphics[totalheight=4.5cm,width=6.5cm]{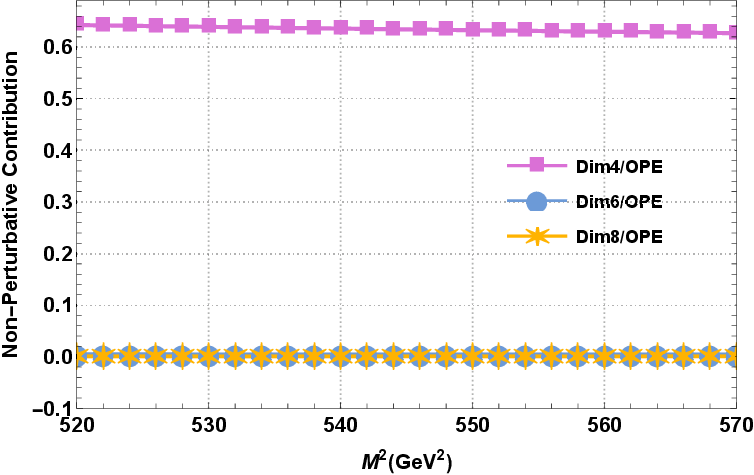}
		\includegraphics[totalheight=4.5cm,width=6.5cm]{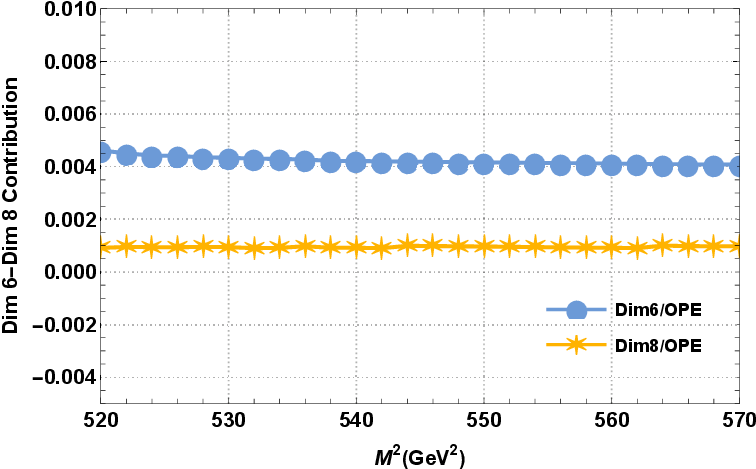}
					\end{center}
	\caption{\textbf{Left:}  Dependence of OPE terms contributions ($ D=4, 6, 8 $)  for the  triply-top baryon ($\Omega_{ttt}$) on the Borel mass parameter ($M^2$) at thecentral value  of the continuum threshold ($s_0$). \textbf{Right :} Dependence of OPE terms contributions ($ D=6, 8 $)  for the  triply-top baryon ($\Omega_{ttt}$) on the Borel mass parameter ($M^2$) at thecentral value  of the continuum threshold ($s_0$).}
	\label{gr11}
\end{figure}

	The continuum threshold $s_0$, which accounts not only for a mathematical cutoff but also for the physical contributions of excited and continuum states, is chosen to ensure the dominance of the ground state within the working window. The selected $s_0$ values for the studied particles are listed in Table \ref{tab:example}.
 The pole dominance and the convergence of the OPE expansion are two additional standard criteria within the QCD sum rule framework. These conditions, respectively, determine the upper and lower limits of the next auxiliary parameter $M^2$, the Borel parameter. Technically, these conditions are satisfied through the relations presented below:
 
 \begin{eqnarray}
\mathrm{PC}=\frac{\Pi(s_0,M^2)}{\Pi(\infty,M^2)}\geq 0.5
\end{eqnarray}
and
\begin{equation}
	 \frac{\Pi ^{\mathrm{Dim8}}(s_0,M^2)}{\Pi (s_0,M^2)}\le\ 0.05,
	  	\label{eq:Convergence}
	  \end{equation}  	
where $\mathrm{PC} $ denotes the pole contribution. The working intervals of the Borel parameter $M^2$ for the studied states are also summarized in Table \ref{tab:example}. While the stability criteria mentioned above are satisfied for all channels, our numerical analysis reveals a characteristic structural pattern in the relative magnitudes of the OPE terms. In heavy-quark systems, the large top-quark mass establishes the dominant physical scale. This significantly enhances the correlator’s sensitivity to non-perturbative vacuum fields. The correlation function, expressed via propagators, generates non-perturbative terms across various mass dimensions by Eq.~(\ref{eqQCD2}). Within the working Borel window, the substantial top-quark mass,  which appears explicitly in the correlation function, enhances the gluon condensate contributions. Consequently, these contributions achieve numerical significance without compromising the convergence of the OPE.  In the baryonic channel $\Omega_{ttt}$, the nonperturbative contribution of dimension $D=4$ (the gluon condensate), amounting to $63.7\%$, dominates over the perturbative term ($D=0$), which contributes $35.8\%$. Similarly, in the mesonic (toponium) channels, a comparable anomaly arises at dimension $D=6$, whose contribution increases to $61.2\%$, surpassing the perturbative part. In light of this observed pattern, the convergence of the OPE remains valid—since the dimension $D=8$ contributions are negligible (about $0.09\%$ in both cases)—and the stability and consistency of the extracted results are ensured through the analysis of their dependence on $M^2$. To demonstrate the convergence of the OPE terms, Fig. \ref{gr11} has been generated for triply- top baryon as an example. The left panel of Fig. \ref{gr11} displays the relative contributions of successive OPE terms with dimensions $D=4, 6, 8$ as a function of the Borel parameter (at the central value of $s_0$). The right panel provides a magnified view of the contributions from terms with $D=6, 8$
, highlighting the specific impact of each of these operators on the convergence.

\begin{figure}[h!]
	\begin{center}
		\includegraphics[totalheight=4.5cm,width=6.5cm]{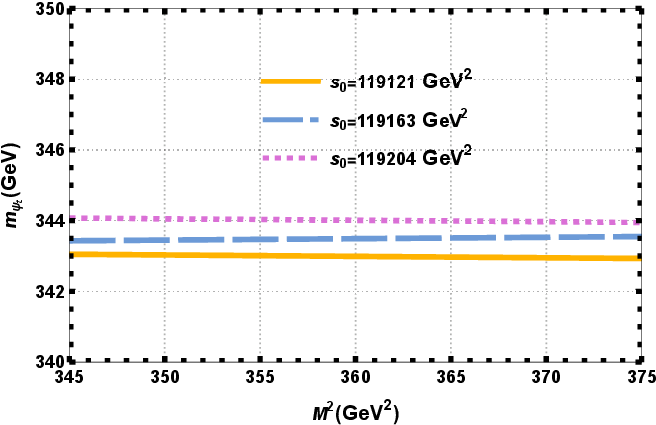}
		\includegraphics[totalheight=4.5cm,width=6.5cm]{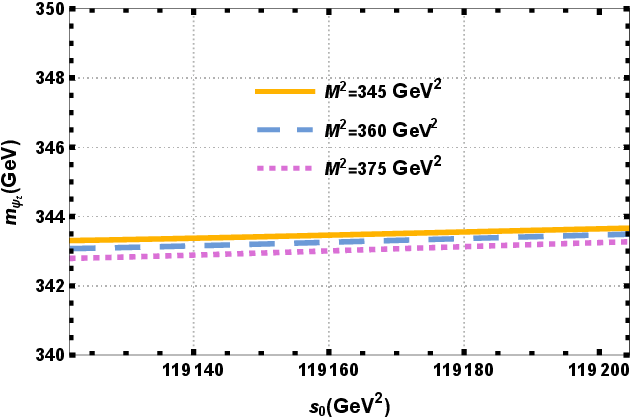}
		\includegraphics[totalheight=4.5cm,width=6.5cm]{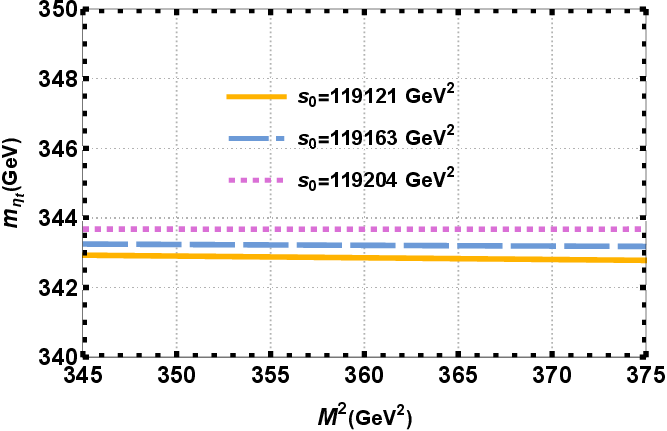}
		\includegraphics[totalheight=4.5cm,width=6.5cm]{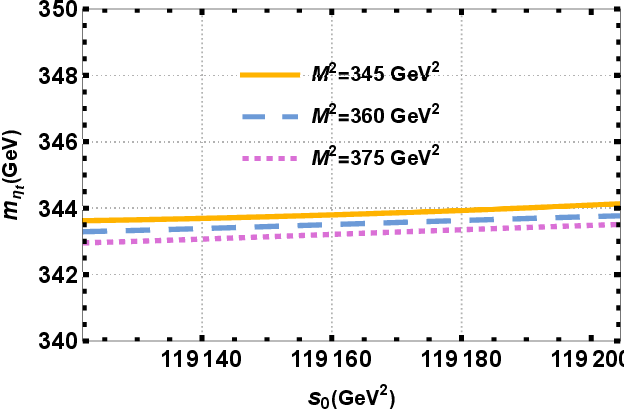}
		\includegraphics[totalheight=4.5cm,width=6.5cm]{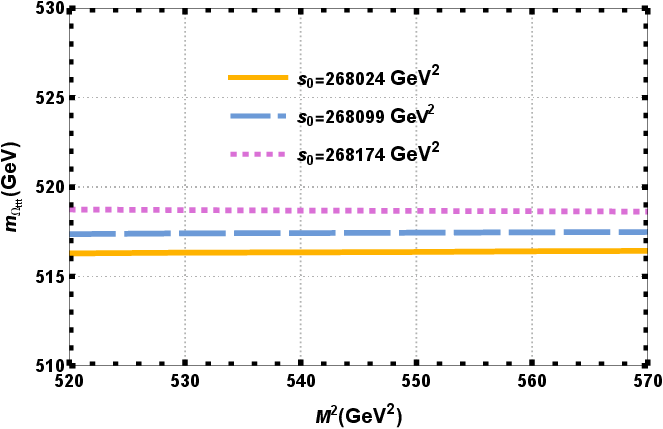}
		\includegraphics[totalheight=4.5cm,width=6.5cm]{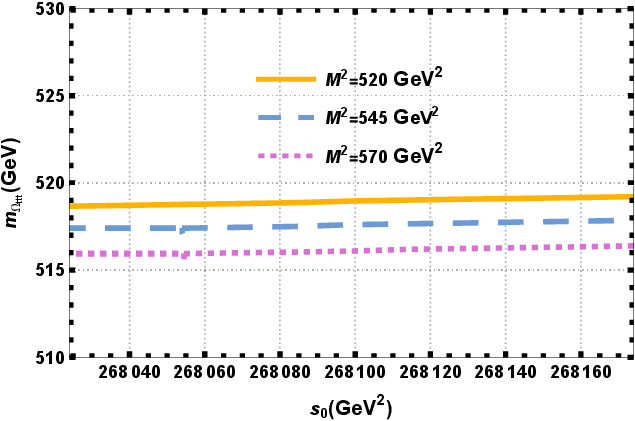}
					\end{center}
	\caption{\textbf{Left panel:}  Dependence of the masses of the pseudoscalar toponium ($\eta_t$), vector toponium ($\psi_t$), and triply-top baryon ($\Omega_{ttt}$) on the Borel mass parameter ($M^2$) for several fixed values of the continuum threshold ($s_0$). \textbf{Right  panel:} Dependence of the masses of the same states on the continuum threshold ($s_0$) for various fixed values of the Borel mass parameter ($M^2$).}
	\label{gr3}
\end{figure}	
	
\begin{table}[ht!]
\centering
\renewcommand{\arraystretch}{0.85} 
\setlength{\tabcolsep}{6pt}       
\fontsize{9}{11}\selectfont
\begin{tabular}{|c|c|c|c|}
\hline
Particle& $M^2~(\mathrm{GeV^2})$&$s_0(\mathrm{GeV^2})$  & m $(\mathrm{GeV})$ \\
\hline
$\eta_t$ & 345-375 & 119121$-$119204 & $343.53{}^{+1.19}_{-1.31}$ \\ \hline
$\psi_t$ & 345-375 & 119121$-$119204 & $343.59{}^{+1.17}_{-1.28}$ \\ \hline
$\Omega_{ttt}$ &520 -570& 268024$-$268174&  $517.81{}^{+1.82}_{-1.88}$\\
\hline
\end{tabular}
\caption{QCD sum-rule determination of the masses of the pseudoscalar  and vector toponium, and the triply-top baryon, within the respective Borel and continuum stability windows. 
}
\label{tab:example}
\end{table}

Fig. \ref{gr3} illustrates the stability of the predicted masses of $\eta_{t}$, $\psi_{t}$, and $\Omega_{ttt}$ with respect to variations in the parameters $s_0$ and $M^2$. As can be clearly seen, the extracted masses exhibit minimal dependence on these parameters within the chosen working windows, confirming the robustness and reliability of the QCD sum-rule results. After determining the optimal working intervals of the auxiliary parameters, the calculated mass values for $\eta_{t}$, $\psi_{t}$, and $\Omega_{ttt}$ are presented in Table \ref{tab:example}. The quoted uncertainties are obtained by combining the errors arising from the sum-rule procedure (including the quark–hadron duality assumption), the variation of the Borel parameter $M^2$ and continuum threshold $s_0$, as well as the uncertainties of the input parameters such as the heavy-quark mass and nonperturbative condensates.	
As seen from Table \ref{tab:example}, the masses of both pseudoscalar and vector toponium states are smaller than the sum of the masses of their constituent top and anti-top quarks. For these bound configurations, the binding energies are given by
\begin{align}
E_{b(\eta_t)} &= 343.53{}^{+1.19}_{-1.31} - 345.12{}^{+0.62}_{-0.62}=-1.59{}^{+0.57}_{-0.69}~\mathrm{GeV}, \nonumber\\
E_{b(\psi_t)} &= 343.59{}^{+1.17}_{-1.28} - 345.12{}^{+0.62}_{-0.62}=-1.53{}^{+0.55}_{-0.66}~\mathrm{GeV},
\end{align}
respectively. This mass deficit may reflect the presence of strong quantum correlations within the $t\bar{t}$ system. In contrast to the two-body toponium states, the triply-top baryon ($\Omega_{ttt}$) is a three-quark system that must satisfy the color-singlet constraint, stabilized by strong inter-quark interactions. While the calculated central mass, $m_{\Omega_{ttt}} = 517.81{}^{+1.82}_{-1.88}\,\mathrm{GeV}$, slightly exceeds the sum of the constituent top-quark masses, $\sum m_t = 517.68{}^{+0.93}_{-0.93}\,\mathrm{GeV}$, the lower bound of the uncertainty range falls below this threshold. This behavior should not be interpreted as a physically negative binding energy, but rather as a reflection of the intrinsic uncertainties of the QCD sum-rule framework, including variations of the Borel mass and continuum threshold parameters.
From a qualitative standpoint, one may speculate that quantum correlations  among the three top quarks could play a role in maintaining the internal colorless configuration and stabilizing the baryonic state, although a more detailed theoretical and experimental investigation is required to clarify these effects. . The CMS collaboration has recently reported an excess of events near the $t\bar{t}$ threshold in final states with two charged leptons and multiple jets, using proton–proton collision data at $\sqrt{s}=13~\mathrm{TeV}$ corresponding to an integrated luminosity of 138 fb$^{-1}$ \cite{CMS:2025kzt}. This excess, observed with a statistical significance exceeding $5\sigma$, was interpreted as evidence for a quasi-bound pseudoscalar toponium state (${{}^1\mathrm{S}_0^{[1]}}$). The measured cross section relative to the fixed-order pQCD prediction is $8.8^{+1.2}_{-1.4}$ pb. No direct mass measurement of the pseudoscalar toponium was provided, though these results offer valuable guidance for theoretical studies. In parallel with the CMS results, the ATLAS Collaboration performed an independent and complementary analysis of $t\bar{t}$ production near the kinematic threshold using 140 fb$^{-1}$ of proton–proton collision data at $\sqrt{s}=13$ TeV.  The study focused on dileptonic final states with multiple jets and compared the observed invariant-mass spectrum to both a baseline perturbative QCD prediction and an extended model incorporating simulations of colour-singlet quasi-bound state formation. A statistically significant excess of events was observed over the baseline expectation, with a significance of 7.7 standard deviations, consistent with the production of an S-wave colour-singlet quasi-bound $t\bar{t}$ state as predicted by non-relativistic QCD. The corresponding excess cross section was measured to be $9.0 \pm 1.3$ pb, further supporting the interpretation of near-threshold dynamics beyond pure perturbative descriptions.
To facilitate a comprehensive comparison in light of recent $t\bar{t}$ measurements, Table \ref{comparison_results} presents the mass estimates for $\eta_t$, $\psi_t$, and $\Omega_{ttt}$ from various  theoretical studies, demonstrating notable consistency within the reported uncertainties. This overall agreement further reinforces the reliability of the present analysis.
\begin{table}[h!]
\centering
\renewcommand{\arraystretch}{1.2}
\begin{tabular}{|c|c|c|c|}
\hline
\textbf{Method}  & \(\mathbf{m(\eta_t)}\)$(\mathrm{GeV})$  & \(\mathbf{m(\psi_t)}\) $(\mathrm{GeV})$& \(\mathbf{m(\Omega_{ttt})}\)$(\mathrm{GeV})$\\
\hline
Present work &$343.53{}^{+1.19}_{-1.31}$ &$343.59{}^{+1.17}_{-1.28}$&$517.81{}^{+1.82}_{-1.88}$ \\
\hline
Non-relativistic potential model \cite{Zhu:2025ezg} & - &-&$513.58 \pm 0.87 \pm 0.23$    \\
\hline
Relativistic potential model \cite{Luo:2025psq} &  $343.29 $ & $343.29 $ &-\\
\hline
Non-relativistic potential model \cite{Jiang:2024fyw} &  $341.26 $ & $341.65 $ &-\\
\hline
\end{tabular}
\caption{Masses of the pseudoscalar ($\eta_t$) and vector ($\psi_t$) toponium, and the triply-top baryon ($\Omega_{ttt}$), compared with  theoretical predictions from various approaches.}
\label{comparison_results}
\end{table}

\section {Conclusion}\label{sec:four}		
In this study, the masses of heavy hadrons containing the top quark—including the pseudoscalar toponium ($\eta_{t}$), vector toponium ($\psi_{t}$), and the triply-top baryon ($\Omega_{ttt}$)—have been investigated within the framework of QCD sum rules. The obtained mass results for $\eta_{t}$ and $\psi_{t}$ are consistent with existing theoretical predictions and phenomenological estimates, and in the case of $\eta_s$ aligns well with the constraints suggested by recent experimental observations.  Another notable outcome of this work is the prediction of the mass of the triply-top baryon $\Omega_{ttt}$, representing the first precise QCD sum-rule determination for this ultraheavy system. Furthermore, our numerical analysis of the OPE indicates a characteristic pattern in the relative contributions of different terms in heavy-quark systems. In particular, the dimension $D=4$ nonperturbative term dominates in the baryonic channel $\Omega_{ttt}$, while in the toponium channels a similar dominance is observed for the dimension $D=6$ term. This observation highlights the notable role of nonperturbative effects at very high quark-mass scales.

Beyond the mass determinations, the present results provide insights into the possible quantum structure of these ultra heavy top-quark bound states. In the two-body toponium states, the observed mass deficit relative to the free-quark threshold may reflect underlying spin–spin correlations between the top and anti-top quarks, which could contribute to the stability of these bound configurations. In contrast, in the triply-top baryon ($\Omega_{ttt}$), quantum correlations are expected to play a qualitatively different role: multi-body correlations among the three top quarks may be relevant for maintaining a color-singlet configuration and the internal consistency of the baryonic state. Overall, these observations highlight the potential relevance of quantum correlations as an intrinsic feature of top-quark systems, while further theoretical and experimental studies are required to clarify their quantitative impact. The results obtained in this study — particularly the predicted masses, such as that of $\Omega_{ttt}$ — serve as precise and valuable theoretical inputs for future experimental searches for these unusual possible bound states in high-energy colliders.

\section*{APPENDIX: Analytical Formulation on the QCD Side of the Sum Rules }
The explicit analytical expressions for both the perturbative and nonperturbative contributions  on the QCD side, corresponding to the pseudoscalar channel ($\eta_t$), are presented below:
\begin{eqnarray} 
		\rho^{\mathrm{pert(PS)}}(s)&=&\frac{-3}{8 \pi ^4}
		\int^{1}_{0} dz  \, \Theta(D)\, e^{\frac{-s}{M^2}} \,m_t^2.
	\end{eqnarray} 	
	
	\begin{eqnarray} 
		\Gamma^{\mathrm{Dim4(PS)}}(M^2)&=&
		\int^{1}_{0} dz   \Big\langle\frac{\alpha_{s}GG}{\pi}\Big\rangle\ e^{-\frac{K}{M^2}}\, \frac{\pi}{L^2\,z^3}\Big[ \frac{m_t^2}{2 M^2 }+\frac{m_t^4}{3 M^4 \, z} \Big].
				    	\end{eqnarray} 
					
\begin{eqnarray} 
		\Gamma^{\mathrm{Dim6(PS)}}(M^2)&=&
		\int^{1}_{0} dz   \Big\langle\ g_s^3 G^3\Big\rangle\ \, \frac{e^{-\frac{K}{M^2}}}{8\pi^2 \,M^2\,L^5\,z}\Big[ (9 - 15 z + 4 z^2)-\frac{m_t^2\,9 (2 - 4 z + 11 z^3 - 12 z^4 + z^5)}{8\, M^2 \,L\, z^4} \nonumber\\
		&+& \frac{m_t^6 (-3 + z + 12 z^2 - 14 z^3 + 5 z^5)}{4\, M^6\,L^3\, z^6 }-\frac{m_t^4 (-12 + 14 z + 20 z^2 - 40 z^3 + 9 z^4 + 15 z^5)}{2 M^4 \,L^2\, z^5}\Big].
\end{eqnarray} 						
	\begin{eqnarray} 
		\Gamma^{\mathrm{Dim8(PS)}}(M^2)&=&
		\int^{1}_{0} dz   \Big\langle\frac{\alpha_{s}GG}{\pi}\Big\rangle\ ^2 e^{-\frac{K}{M^2}}\, \frac{\pi^2\,m_t^2}{3 M^6\,L^5\,z^5}\Big[ \frac{m_t^4}{4 M^4\, L\,z }+\frac{m_t^2(-1 + 5 z + 8 z^2)}{2 M^2\,L \, z} \nonumber\\
		&+& (2 - 5 z - 11 z^2) \Big].
				    	\end{eqnarray} 	
		Here, $\Theta(D)$ denotes the unit step function, and the following shorthand notations are employed:				
				\begin{eqnarray}
	D &=&-m_t^2 - s\, L\, z,\nonumber\\
	K &=&-\frac{m_t^2}{L\,z},\nonumber\\
		L&=&z-1.
\end{eqnarray}	
							

\end{document}